\documentclass[conference]{IEEEtran}
\IEEEoverridecommandlockouts

\usepackage{cite}
\usepackage{amsmath,amssymb,amsfonts}
\usepackage{algorithmic}
\usepackage{graphicx}
\usepackage{textcomp}
\usepackage{xcolor}
\usepackage{url}
\usepackage{eurosym}

\def\BibTeX{{\rm B\kern-.05em{\sc i\kern-.025em b}\kern-.08emT\kern-.1667em\lower.7ex\hbox{E}\kern-.125emX}}

\usepackage[top=0.75in,bottom=1in,left=0.68in,right=0.68in]{geometry}

\newcommand{\Setup}{{\sf Setup}}
\newcommand{\OKGen}{{\sf OKGen}}
\newcommand{\UKGen}{{\sf UKGen}}
\newcommand{\Sign}{{\sf Sign}}
\newcommand{\Verify}{{\sf Verify}}
\newcommand{\Open}{{\sf Open}}
\newcommand{\Judge}{{\sf Judge}}
\newcommand{\pp}{\mathsf{pp}}
\newcommand{\opk}{\mathsf{opk}}
\newcommand{\osk}{\mathsf{osk}}
\newcommand{\pk}{\mathsf{pk}}
\newcommand{\sk}{\mathsf{sk}}
\newcommand{\dec}{\mathsf{dec}}

\newtheorem{definition}{Definition}

\begin{document}
\title{An Anonymous Trust-Marking Scheme on Blockchain Systems\thanks{$\ast$~An extended abstract appeared at the 3rd IEEE International Conference on Blockchain and Cryptocurrency, ICBC 2021. This is the full version. }}

\date{\today}


\author{\IEEEauthorblockN{Teppei Sato}
\IEEEauthorblockA{\textit{University of Tsukuba}\\\textit{Japan}}\\
\and
\IEEEauthorblockN{Keita Emura}
\IEEEauthorblockA{\textit{National Institute of}\\\textit{Information and}\\\textit{Communications Technology}\\\textit{Japan}}\
\and 
\IEEEauthorblockN{Tomoki Fujitani}
\IEEEauthorblockA{\textit{University of Tsukuba}\\\textit{National Institute of}\\\textit{Information and}\\\textit{Communications Technology}\\\textit{Japan}}\\
\and
\IEEEauthorblockN{Kazumasa Omote}
\IEEEauthorblockA{\textit{University of Tsukuba}\\\textit{National Institute of}\\\textit{Information and}\\\textit{Communications Technology}\\\textit{Japan}}\\
}

\maketitle

\begin{abstract}
During the Coincheck incident, which recorded the largest damages in cryptocurrency history in 2018, it was demonstrated that using Mosaic token can have a certain effect. 
Although it seems attractive to employ tokens as countermeasures for cryptocurrency leakage, Mosaic is a specific token for the New Economy Movement (NEM) cryptocurrency and is not employed for other blockchain systems or cryptocurrencies. 
Moreover, although some volunteers tracked leaked NEM using Mosaic in the CoinCheck incident, it would be better to verify that the volunteers can be trusted. Simultaneously, if someone (e.g., who stole cryptocurrencies) can identify the volunteers, then that person or organization may be targets of them. 
In this paper, we propose an anonymous trust-marking scheme on blockchain systems that is universally applicable to any cryptocurrency. 
In our scheme, entities called token admitters are allowed to generate tokens adding trustworthiness or untrustworthiness to addresses. Anyone can anonymously verify whether these tokens were issued by a token admitter. 
Simultaneously, only the designated auditor and no one else, including nondesignated auditors, can identify the token admitters. 
Our scheme is based on accountable ring signatures and commitment, and is implemented on an elliptic curve called Curve25519, and we confirm that both cryptographic tools are efficient. Moreover, we also confirm  that our scheme is applicable to Bitcoin, Ethereum, and NEM. 
\end{abstract}

\section{Introduction}

Cryptocurrencies (cryptoassets) such as Bitcoin, Ethereum, and NEM have attracted attention in various aspects, such as the large market capitalization and the use of smart contracts. Because of a cryptocurrency's decentralized structure, there is merit in fault tolerance and anyone can check the content of transactions; thus, a certain level of transparency is guaranteed. However, cryptocurrencies have been used for crimes because of their anonymity and have become a target of cyberattacks as their value increases. Furthermore, since it seems difficult to stop suspicious transactions such as unauthorized withdrawals or exchanges before they happen, it is important to consider countermeasures after cryptocurrencies are stolen.

During the Coincheck incident, which recorded the largest damages in cryptocurrency history in 2018, some volunteers tracked leaked New Economy Movement (NEM) cryptocurrencies using Mosaic, which is a NEM-specific feature allowing a self-made token to be sent on the NEM blockchain. These volunteers sent Mosaic to addresses that received leaked NEMs from Coincheck addresses and warned the recipients that these should not be exchanged for other cryptocurrencies or legal currencies. A criminal can discard such marked addresses, so it can be seen as a cat-and-mouse game. It was not recognized that a large amount of NEM was exchanged via legally authorized exchanges. The criminal finally exchanged stolen NEM via a self-made exchange established on the dark web (hidden service) until the removal of the system on March 18, 2018~\cite{nemio_announcement1}. Thus, the effectiveness of tokens in preventing the spread of leaked NEMs is admitted. Because theft of cryptocurrencies has become a real risk, it seems effective to employ tokens as countermeasures for cryptocurrency leakage. However, there are issues to be solved:

\begin{itemize}
\item Mosaic is a specific token for NEM, and it is not employed for other blockchain systems or cryptocurrencies.

\item There is no way to check whether the volunteers can be trusted.  

\item If someone (e.g., who stole cryptocurrencies) can identify the volunteers, then that person or organization may be targets of them. 
\end{itemize}

\noindent
From a functionality standpoint, it seems effective to add trustworthiness to addresses, but the Mosaic token instead adds untrustworthiness to addresses. For example, if an address is authorized by adding a token for trustworthiness, then anyone can recognize it by checking whether the token is included in the corresponding transaction.

One might think that claims, which are defined in the ERC-725/ERC-735 Ethereum Identity Standard\footnote{See \url{https://erc725alliance.org/} or \url{https://github.com/ethereum/EIPs/issues/735}.} could be used to add and remove tokens. When a user wants to deploy smart contracts, a claim issuer (who may or may not be the user) can add some information relative to the user's identity, and it can be verified as a credential. However, anonymity of claim issuers is not a primary subject of the standards; thus, it cannot be directly employed for our usage. 
One naive solution is to employ anonymous signatures such as ring signatures~\cite{rivest2001leak}, especially linkable ring signatures~\cite{LiuWW04,au2006short,TorresSSLKBAC18,SunALY17,WangCM19,LuAZ19}, which have been widely used to add anonymity in cryptocurrencies such as Monero~\cite{Noether15}. This solution assumes that there are token admitters who are trusted; i.e., they are recognized as entities that can issue tokens but are not allowed further responsibilities. They generate a signature and include the signature on a transaction. In the signed message, anyone can check whether the address is authorized or unauthorized by verifying the signature. This method is applicable to any blockchain system. Moreover, anyone can anonymously check whether a token admitter generates a signature. This naive solution looks promising, but an issue of anonymity arises. For example, if an address will behave maliciously using a trustworthiness token, then the trustworthiness of the corresponding token admitter should be reduced. Moreover, if some users who have addresses deny the addition of a token for untrustworthiness, then the corresponding token admitter needs to be traced. However, there is no way to identify the corresponding token admitter if (linkable) ring signatures are employed. Thus, it seems effective to consider auditors, who can identify token admitters from signatures. 

Another naive solution is to employ group signatures~\cite{ChaumH91}. In this scenario, a group manager issues signing keys to users. A user generates a signature, and a verifier verifies the signature anonymously, i.e., checks whether the signer is a group member. If an accident happens, the group manager can identify the signer using a secret key. However, this centralized structure does not match decentralized blockchain systems. If we consider permissioned blockchain systems, then there is an additional access control layer and it may allow us to prepare such centralized organizations. 
However, it would be better to cover both public and permissioned blockchain systems. 

\medskip
\noindent\textbf{Our Contribution}: In this paper, we propose a trust-marking scheme on blockchain systems explained as follows: 

\begin{itemize}

\item The trust-marking scheme is universally applicable to any cryptocurrency, and works even in public blockchains. 
 
\item Token admitters can anonymously issue tokens for either trustworthiness or untrustworthiness to addresses by generating accountable ring signatures~\cite{XuY04,bootle2015short,KumawatP17,LaiZCS16}. 

\item Token admitters designate an auditor when they issue tokens, and only the auditor can identify who issued the tokens.

\end{itemize}

\noindent 
As building blocks, we employ accountable ring signatures~\cite{XuY04,bootle2015short,KumawatP17,LaiZCS16}, which can be seen as a generalization of ring and group signatures (See Section~\ref{ars}). To our knowledge, an attempt to employ accountable ring signatures as tokens has not previously been made, although ring signatures have been widely employed in blockchain systems. By checking accountable ring signatures, anyone could anonymously check whether a token admitter generated a signature. Moreover, because of anonymity, even if someone who stole cryptocurrencies wanted to retaliate against the corresponding token admitter, that person or organization would not be able to identify the token admitter. If an accident happened, only the designated auditor could identify the signer. In other words, our scheme preserves decentralized structures, because nondesignated auditors cannot break anonymity. We also consider a case in which a token admitter revokes trustworthiness or untrustworthiness tokens. Subsequently, only the actual token admitter can issue a trustworthiness or untrustworthiness token. The fact that two tokens were issued by the same token admitter is revealed but still the token admitter is anonymous.


We implemented our scheme using the Bootle et al. accountable ring signature scheme~\cite{bootle2015short} and the Pedersen commitment scheme~\cite{Pedersen91}, employed Curve25519~\cite{Bernstein06}. 
To confirm whether our scheme is universally applicable to any cryptocurrency, we selected Bitcoin, Ethereum, and NEM for capturing various features.\footnote{Briefly, Bitcoin adopts UTXO (unspent transaction output) while Ethereum and NEM adopt account based transaction models. Bitcoin and Ethereum adopt PoW (Proof of Work) while NEM adopts PoI (Proof of Importance). Our scheme is applicable regardless of these features.} 

\medskip
\noindent\textbf{Related Work}: The word \lq\lq Tokens'' has been widely used in other blockchain systems.  Here, for the sake of clarity and for explaining the difference from our work, we introduce these works as follows. 
Basically, a token is a non-sensitive surrogate value which is replaced from sensitive elements (this replacement procedure is so-called tokenization), and is employed for privacy-preserving payment system. Here, token is regarded as cryptocurrency itself. 
To name a few: Zerocoin~\cite{MiersG0R13}, Zerocash~\cite{Ben-SassonCG0MTV14}, Confidential Assets~\cite{PoelstraBFMW18}, QuisQuis~\cite{FauziMMO19}, and Zether~\cite{BunzAZB20}. 

For adding auditability (as in our scheme), Androulaki et al.~\cite{AndroulakiCCDET20} proposed a privacy-preserving and auditable token management system for permissioned blockchain systems such as Hyperledger Fabric~\cite{AndroulakiBBCCC18}. 
The Androulaki et al. system architecture is somewhat similar to our scheme where users, issuers, and auditors are defined (in their paper a certifier and a registration authority are also defined). 
Briefly, users own tokens that represent some real-world assets. They wish to exchange their tokens with other users in the network. Issuers are authorized to introduce tokens (as in token admitters in our scheme). 
Auditors inspect transactions of users. 
One big difference from our work is that they did not consider anonymity of issuers. Moreover, in their system, auditors are not designated at each token generation phase where each user is assigned an auditor at registration time and that this assignment is basically immutable (updating a user's auditor is possible by updating that user's attribute credential). 
Yuen~\cite{Yuen19} also proposed a private, authenticated and auditable consortium (i.e., partly private) blockchain. The system considers sender/recipient privacy, where the identity of a sender/recipient of a transaction is not revealed to any third party, and transaction privacy, where the content of the transaction is not revealed to any third party. 
In addition, auditability is introduced, where the auditor can recover the sender identity, recipient identity and/or the transaction amount. That is, the audit target is different from our purpose. 
K{\"{u}}sters et al.~\cite{KustersRS20} also considered accountability in permissioned blockchains. They insisted that all parties know each other, and hence, accountability incentivizes all parties to behave honestly. 
We would like to insist that accountability is also important in public blockchains, and we mainly consider this setting as mentioned above. 
Garman et al.~\cite{Garman0M16} proposed a decentralized anonymous payment system in the public blockchain. The system is capable of enforcing compliance with specific transaction policies. They considered a coin tracing scheme where individual coins can be marked for tracing. Briefly, all of the information needed to trace the output coins is encrypted under the existing key for the input coin commitments, and all of this output is encrypted by authority's public key. 
One big difference from our work is that they did not consider both trustworthiness and untrustworthiness. 

Camenisch et al.~\cite{CamenischDD17} proposed a delegatable anonymous credential system. An (attribute-based) anonymous credential is a set of attributes certified to a user by an issuer. A user can prove the possession of a credential via a zero-knowledge proof system by creating a fresh token. Then, the user can select which attributes need to be included when the token is created. 
Because credentials are typically issued in a hierarchical manner, Camenisch et al. considered delegatability of credentials to hide information revealed from the chain of issuers. Moreover, they consider an application of the system to a permissioned blockchain. Basically, in the system users spontaneously generate tokens to prove the possession of a credential. On the other hand, in our scheme users (addresses) are forcibly evaluated and issued a token. 

Zhang et al.~\cite{ZhangLLYAW19} proposed a system that signs cryptocurrency transactions by linkable group signatures~\cite{NakanishiFW1999}. The system enables us to trace a payer's identity in consortium blockchain based anonymous cryptocurrencies in case a payer misbehaves in the system. The anonymity can be retained if a user behaves honestly. 

Token-based traceability systems also have been considered. 
For example, Westerkamp et al.~\cite{WesterkampVK18} proposed e a blockchain-based supply chain traceability system using smart contracts. 
A non-fungible token (NFT), which is an identifiable token reflecting digital goods or physical goods, is standardized in ERC-721\footnote{See \url{https://github.com/ethereum/EIPs/blob/master/EIPS/eip-721.md}.}, corresponds to a batch of physical goods. 
Watanabe et al.~\cite{WatanabeIO0NHK19} proposed a token-based traceability system based on a directed acyclic graph (DAG). They insisted that NFT is expected to be able to represent complicated operations in traceability systems, everyone needs to easily confirm the history of the circulation of tokens related to each product. Their system allows us to efficiently retrieval of past histories. 
In these use cases, tokens are basically used for tracing a product's provenance. 
On the other hand, our scheme evaluates (un)trustworthiness of addresses themselves.

Omote et al.~\cite{OSS20} proposed a system that adds (un)trustworthiness to addresses where a special address (as in a token admitter of our scheme) sends a few Bitcoin to them as a token. As differences from our work, they considered Bitcoin only and did not consider anonymity of the token admitter. 

\section{Preliminaries}
\label{ars}

\noindent\textbf{Accountable Ring Signatures}: 
In blockchain systems, a transaction is signed by a secret key connected with an address. According to applications, anonymous signatures (as the underlying signature scheme) and a decentralized structure are required. Thus, ring signatures~\cite{rivest2001leak} rather than group signatures are employed, especially in the cryptocurrency context. 
The ring signatures enabled a signer to sign anonymously. Each user generated a public verification key and a secret signing key, selected a ring that is a set of verification keys, and generated a signature on behalf of the ring. A verifier could verify whether the signer was a member of the ring. 
Unlike pseudonyms, ring signatures provide unlinkability (i.e., even if the same signer generated two signatures, nobody would know those signatures were generated by the same signer) and unforgeability (defined as in digital signatures) 
(see~\cite{BenderKM09} for the detailed security definition). 
To prevent double spending in privacy-preserving cryptocurrencies, linkable ring signatures~\cite{LiuWW04,TorresSSLKBAC18,SunALY17}, providing weak anonymity compared with that of ring signatures, were employed, e.g., in CryptoNote~\cite{van2013cryptonote} (see \cite{TorresSSLKBAC18} for the detailed security definition of (one-time) linkable ring signatures). 
A ring confidential transaction (RingCT)~\cite{Noether15,YuenSLAEZG20,SunALY17} employed a similar cryptographic technique. 

Although strong anonymity (meaning no entity can identify the actual signer) is an attractive and effective security in a privacy-preserving context, it may cause unexpected incidents (e.g., signers may behave maliciously because they will not be identified). In group signatures, a group manager has the power to track a signer, but this ability is too strong, and the group manager can become a big brother. In particular, its centralized structure does not match the structure of blockchain systems. Xu and Yung~\cite{XuY04} proposed accountable ring signatures as a generalization of ring signatures and group signatures. As in ring signatures, each user generated a public verification key and a secret signing key. Openers generated own public key and secret opening key. A user made a ring (a set of verification keys), designated an opener public key, and generated a signature using his or her own signing key. A verifier verified whether a signer was contained in the ring, and only the designated opener could identify the actual signer. 
Later, Bootle et al. formalized the security definitions. In particular, they considered signature hijacking attacks~\cite{SakaiSEHO12} and provided the function that allowed the opening result to be publicly verified~\cite{BellareSZ05}. Bootle et al. also provided a concrete accountable ring signature scheme with a logarithmic signature size of the number of ring members. 
The Bootle et al. scheme is based on the Camenisch-Stadler group signature scheme~\cite{CamenischS97} and the Groth-Kohlweiss one-out-of-many proofs of knowledge~\cite{GrothK15}. 
It was secure under the decisional Diffie-Hellman (DDH) assumption (in prime-order groups), which is widely recognized as a standard cryptographic assumption. Moreover, assuming the hardness of the DDH problem allows us to employ Curve25519~\cite{Bernstein06} to implement the scheme.
\footnote{Symmetric pairings, defined as a map from points on elliptic curves to an element on a finite field, are known as a DDH solver. It is well known that Curve25519 cannot be used for pairing-based cryptography, i.e., Curve25519 is not a pairing-friendly curve. Thus, we can assume that the DDH assumption holds.} 
In contrast, Lai et al.~\cite{LaiZCS16} showed a relationship between sanitizable signatures and accountable ring signatures and proposed an accountable ring signature scheme with a constant signature size. One downside of the scheme was that it employed composite-order bilinear groups. 
As an alternative, Kumawat et al.~\cite{KumawatP17} proposed another accountable ring signature scheme from indistinguishability obfuscation ($i\mathcal{O}$)~\cite{BarakGIRSVY01} which is recognized as one of the extremely strong cryptographic tools. Thus, we mainly followed the Bootle et al. scheme in this paper. 

\medskip
\begin{definition}[Syntax of Accountable Ring Signatures~\cite{bootle2015short}]~
\begin{itemize}
\item $\Setup(1^\lambda)$: 
The setup algorithm takes as input a security parameter $\lambda\in\mathbb{N}$ and outputs a common parameter $\pp$. 

\item $\OKGen(\pp)$: The opener key generation algorithm takes as input $\pp$, and outputs an opener public key $\opk$ and a secret key $\osk$. 

\item $\UKGen(\pp)$: The user key generation algorithm takes as input $\pp$, and outputs a user public key $\pk$ and a secret key $\sk$. 

\item $\Sign(\opk,M,R,\sk)$: The signing algorithm takes as input $\opk$, a message to be signed $M$, a ring $R$, and $\sk$ and outputs a ring signature $\sigma$. We assume that the corresponding $\pk$ is contained in $R$. 

\item $\Verify(\opk,M,R,\sigma)$: The verification algorithm takes as inputs $\opk$, $M$, $R$, and $\sigma$ and outputs 1 (accept) or 0 (reject). 

\item $\Open(M,R,\sigma,\osk)$: The opening algorithm takes as inputs $M$, $R$, $\sigma$, and $\osk$ and outputs $\pk$ of the signer and its proof $\pi$ or $\bot$ otherwise. 

\item $\Judge(\opk,M,R,\sigma,\pk,\pi)$: The judge algorithm takes as inputs $\opk$, $M$, $R$, $\sigma$, $\pk$, and $\pi$ and outputs 1 (meaning that $\sigma$ is generated by $\sk$ corresponding to $\pk$) or 0 otherwise. If $\sigma$ is not valid or $\pk\not\in R$, then output 0. 
\end{itemize}
\end{definition}

Correctness is defined as a standard mannar such that, if all keys are honestly generated and a signature is honestly generated using them, then the verification of the signature always outputs 1. Beside correctness, Bootle et al. defined four security notions: full unforgeability, full anonymity, traceability, and tracing soundness. Full unforgeability ensures that an adversary who may corrupt all-but-one users and openers cannot produce a forged signature, proof that the signature is valid, and proof with a verification key that the noncorrupted user is accepted by the  $\Judge$ algorithm. 
Full anonymity ensures that an adversary who may have all signing keys (and secret keys of nondesignated openers, as discussed later) cannot distinguish whether two signatures are generated by the same signing key.\footnote{As a similar primitive, Zhang et al.~\cite{ZhangLSKY19} proposed revocable and linkable ring signatures. One big diffirence from accountable ring signatures is linkability, where anyone can link whether two signatures are generated by the same signing key.} 
Traceability ensures that designated openers can always identify the actual signer and can produce valid proof for their identification. Tracing soundness ensures that a signature cannot trace two different signers. See~\cite{bootle2015short} for the formal security definitions.

\medskip
\noindent\textbf{Remark}: 
The Bootle et al. scheme is secure in the common reference model (i.e., a common parameter $\pp$ needs to be honestly generated). We discuss how to generate $\pp$ in our scheme later. 
The syntax and security definitions of accountable ring signatures given in Bootle et al.~\cite{bootle2015short} considered only one opener. However, because of our usage, e.g., a signer designates an opener, it is natural to consider plural openers.\footnote{As a similar primitive, multi-opener group signatures have been proposed~\cite{BenjumeaCLY08,Ghadafi14,Lu0ZL19} where plural openers identify the signer as a threshold manner. Unlike (accountable) ring signatures, users are required to run the join algorithm to obtain signing keys.
} 
According to this modification, we also need to modify the security definition in which nobody except the designated opener can break anonymity; i.e., an adversary of anonymity is allowed to obtain secret keys of nondesignated openers. This modification does not affect security. Let $\ell$ openers be defined. Then, the adversary is allowed to corrupt (i.e., is allowed to know the secret keys of) at most $\ell-1$ openers. In the security proof of anonymity, just guessing the non-corrupted opener is enough, and its probability is at least $1/\ell$. 
Therefore, we assume that there are plural openers in our scheme.

Bootle et al. introduced fully dynamic group signatures~\cite{BootleCCGG16} where users may join and leave at any time. They mentioned that fully dynamic group signatures can be constructed from accountable ring signatures. Thus, other fully dynamic group signature scheme, e.g.,~\cite{0001HS19}, might be employed for our usage. 

\medskip
\noindent\textbf{Commitment}: 
A commitment scheme ${\sf Com}$ consists of three algorithms. The ${\sf ComSetup}$ algorithm takes as input a security parameter $\lambda\in\mathbb{N}$ and outputs a common reference string ${\sf crs}$. The ${\sf Commit}$ algorithm takes as inputs ${\sf crs}$ and a message $M$ to be committed and outputs a commitment $C$ and a decomit $\dec$. 
The ${\sf ComOpen}$ algorithm takes as inputs ${\sf crs}$, $C$, $M$, and $\dec$ and outputs 1 (meaning that $C$ is a commitment of $M$) or 0 otherwise. 
 A commitment scheme provides both hiding (no information of $M$ is revealed from $C$) and binding ($C$ is not opened to two $(M,\dec)\neq (M^\prime,\dec^\prime)$) properties. 
In our implementation, we employ the Pedersen commitment scheme~\cite{Pedersen91} which provides computational binding (under the discrete logarithm (DL) assumption) and perfect hiding. 
We remark that ${\sf crs}$ needs to be generated honestly.
We discuss how to generate ${\sf crs}$ in our scheme later. 

\medskip
\noindent\textbf{Common Reference String}: 
In the Bootle et al. accountable ring signature scheme, $h_1,\ldots,h_n\in\mathbb{G}$, where $\mathbb{G}$ is a group of prime order $p$, and a public key of the ElGamal encryption scheme $ek=g^\tau$, where $\tau\in \mathbb{Z}_p$, need to be honestly generated. That is, nobody knows  $\log_g h_1,\ldots,\log_g h_n$, and $\tau$. 
If there is no central authority, i.e., in permissionless blockchains, we need to consider how to generate them. In the Pedersen scheme, 
the {\sf ComSetup} algorithm outputs ${\sf crs}=(g,h)$ where $g,h\in\mathbb{G}$ are generators. For $M\in\mathbb{Z}_p$, the ${\sf Commit}$ algorithm randomly chooses $\dec\in\mathbb{Z}_p$ and computes $C=g^M h^{\dec}$. The ${\sf ComOpen}$ algorithm outputs 1 if $C=g^M h^{\dec}$ holds and 0 otherwise. 
Here, ${\sf crs}$ needs to be generated honestly, i.e., nobody knows $\log_g h$. 

We assume that the generator $g$ is publicly specified and we employed the generator of Curve25519 defined by the $x$-coordinate is 9~\cite{rfc7748}. 
We employed the multi-CRS setting~\cite{GrothO07}. 
We assumed that a token admitter $i$ chose $\tau_i\in\mathbb{Z}_p$, computed $ek_i=g^\tau_i$, and broadcast $ek_i$ to other token admitters. Simultaneously, the token admitter $i$ proved the knowledge of $\tau_i$ via a zero-knowledge proof system. Finally, $ek$ is defined as $ek=\sum ek_i$. We also assumed that all token admitters did not collude. $h_1,\ldots,h_n$ and $h$ also could be generated similarly.

\section{Proposed Anonymous Trust-Marking Scheme}

In this section, we give our anonymous trust-marking scheme. As the first attempt, we simply added an audit functionality to CryptoNote~\cite{van2013cryptonote}, which provides anonymity in Monero. 
 CryptoNote originally provided an audit functionality in which secret values were revealed outside the blockchain. However, we could not control who would audit it. Thus, we tried to change (linkable) ring signatures on CryptoNote to accountable ring signatures. Then, a signer who signed an address could designate an auditor and include the auditor public key on the signed message. Anyone could check both the transaction validity and the designated auditor. Therefore, it seemed the signature could be seen as a token. However, in anonymous blockchain systems, addresses are anonymous. 
Thus, we needed to additionally prepare a function for proving possession of tokens, which may be implemented by employing zero-knowledge proof systems, but there is no motivation if an untrustworthiness token is issued. 
So, we constructed our scheme on nonanonymous blockchain systems. 

\subsection{Entities and Roles}

In the system, if token admitters have addresses for sending transactions, then no anonymity is provided. One solution was to share a common address among token admitters. However, because the secret key associated with the address was also shared, all token admitters could withdraw remittance fee freely, a situation that was not desirable. Thus, we separated the right to send transactions and the right to generate token. We defined the entities in our scheme as follows:

\begin{itemize}
\item Users who have addresses to be used for sending transactions. 
\item Token admitters who issue trustworthiness or untrustworthiness tokens to users. 
\item Token submitters who send tokens to users on behalf of token admitters. 
\item Token verifiers who check the validity of tokens; in actual scenarios, they may be other users who would like to check the trustworthiness or untrustworthiness of the user who was issued the token or may be currency exchanges that would like to check that the cryptocurrencies were not stolen. 
\item Auditors who identify token admitters if they are designated by token admitters. 
\end{itemize}

\begin{figure*}[t!]
\centering
\includegraphics[bb=0 0 800 540,scale=0.55]{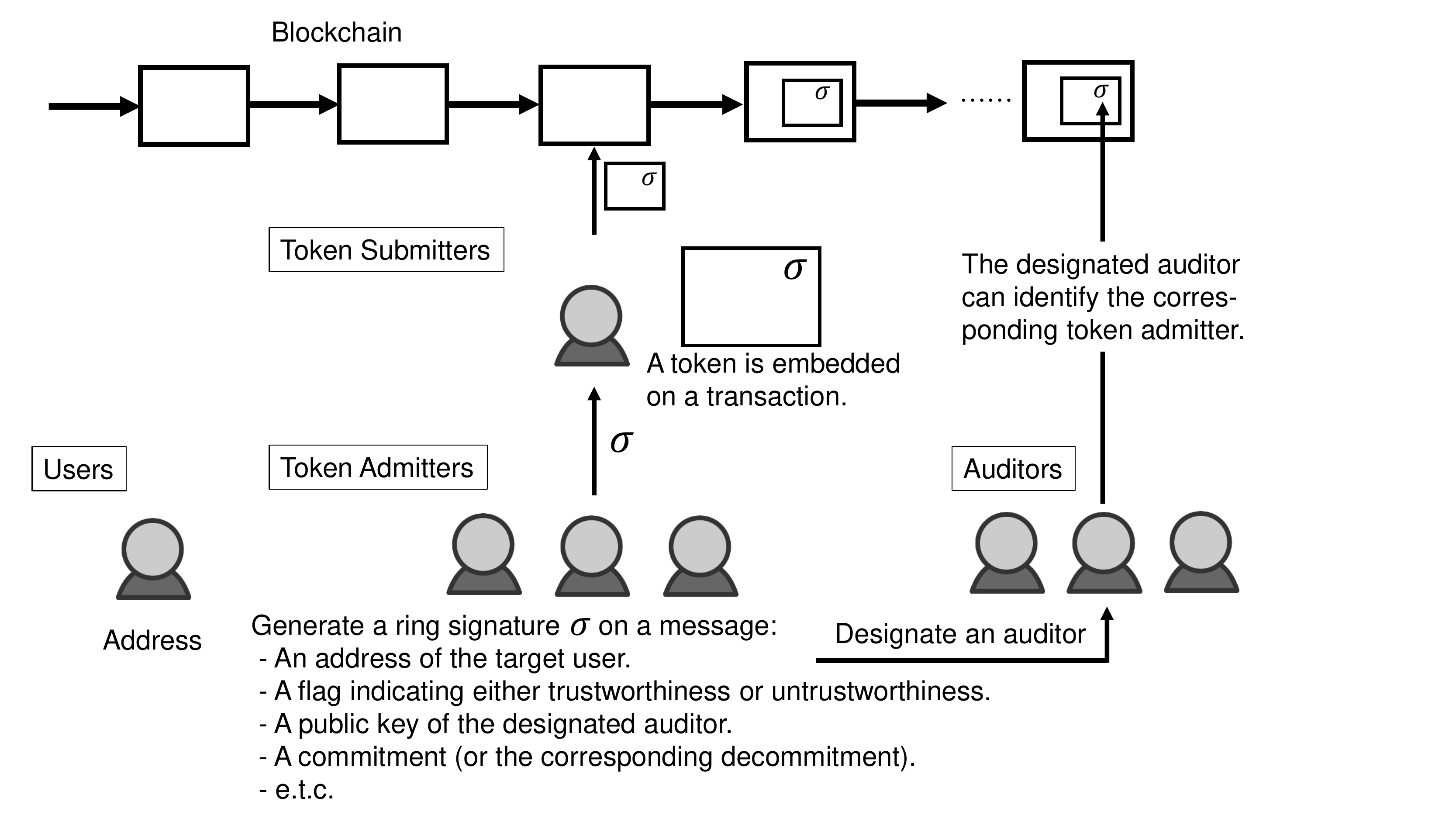}
\caption{Flow of our scheme}\label{flow}
\label{fig1}
\end{figure*}

\noindent 
We also need to consider the case that a token admitter has evaluated the trustworthiness or untrustworthiness of a user (address), and later the trustworthiness or untrustworthiness need to be revoked, e.g., an address behaves maliciously after a token for trustworthiness is issued to the address. This token revocation functionality is also important since once some values are written in a blockchain system, it is quite difficult to remove them from the system owing to the availability of blockchain. We need to guarantee that no token admitter can revoke a token unless the token was issued by themselves. Since token admitters are anonymous, we need to carefully link whether two token admitters are the same without detracting anonymity. We realize it by using a commitment scheme. 

\subsection{Security Requirements}

Basically, we consider four security notions: anonymity, auditability, unforgeability, and revocability. 
These are (informally) defined as follows. 

\begin{itemize}
\item Anonymity: Let a token be generated by a token admitter. Then, nobody, except the designated auditor or the token submitter who sends the token, can identify who the actual token admitter is.  

\item Auditability: No untraced token can be generated by token admitters. Moreover, no designated auditor can cheat the tracing result. 

\item Unforgeability: No address or (un)trustworthiness of a target user specified by a token admitter can be modified. 

\item Revocability: No token admitter can revoke a token unless the token was issued by themselves.  
\end{itemize}

\subsection{Proposed Scheme}

We assume that a common parameter $\pp$ and a common reference string ${\sf crs}$ have been honestly generated. At the initial step, each token admitter runs $(\pk,\sk)\leftarrow \UKGen(\pp)$. We denote $(\pk_i,\sk_i)$ as the $i$-th token admitter. 
Each auditor runs $(\opk,\osk)\leftarrow \OKGen(\pp)$. 
We denote $(\opk_j,\osk_j)$ as the $j$-th auditor. 
Moreover, each token submitter generates an address for sending transactions. 

Next, we define the message format to be signed as follows.

\begin{itemize}
\item Address of the target user.
\item A flag indicating either trustworthiness or untrustworthiness. 
\item Public key of the designated auditor.
\item Expiry time.
\item Either a commitment or both the corresponding decommitment and the hash value of the corresponding transaction (which are explained later). 
\item Any message, e.g., \lq\lq This user appropriately sent transactions.'' or \lq\lq This user received cryptocurrencies *** that were leaked on CCYYMMDD from ***."
\end{itemize}

We show the flow of our scheme in Fig.~\ref{fig1}. The case in which a token admitter issues a token is as follows:

\smallskip
\noindent Token Admitter $i$:
\begin{enumerate}

\item Randomly chooses $r_{\sf link}$ and runs $(C,\dec)\leftarrow {\sf Commit}({\sf crs},r_{\sf link})$. 

\item Defines ring $R$ where $\pk_i\in R$. 

\item Designates an auditor $j$ and let $\opk_j$ be the public key. 

\item According to the message format, prepares a message $M$ to be signed. $M$ contains $C$ and $\opk_j$. 

\item Runs $\sigma\leftarrow \Sign(\opk_j,M,R,\sk_i)$ and sends $(M,\sigma)$ to a token submitter. 

\end{enumerate}

\noindent The case in which a token admitter revokes a token is as follows. Assume that the token admitter has issued a token $\sigma^\prime$ on $M$ on behalf of $R$, that $M$ contains a commitment $C$, and the token admitter knows the corresponding $(r_{\sf link},\dec)$. In other word, no token admitter can revoke a token unless the token was issued by themselves. 

\smallskip
\noindent Token Admitter $i$:
\begin{enumerate}

\item According to the message format, prepares a message $M$ to be signed. $M$ contains $(r_{\sf link},\dec)$ and the hash value of the transaction that $\sigma^\prime$ is embedded. 

\item Runs $\sigma\leftarrow \Sign(\opk_j,M,R,\sk_i)$ and sends $(M,\sigma)$ to a token submitter. 

\end{enumerate}

\noindent Next, the procedure for a token submitter is as follows:

\smallskip
\noindent Token Submitter:
\begin{enumerate}
\item If $R$ contains unknown public key, then abort. 

\item If $\opk_j$ is not a public key of auditors, then abort. 

\item If the expiry time contained in $M$ has passed, then abort. 

\item If $\Verify(\opk_j,M,R,\sigma)=0$, then abort. 

\item Otherwise, embed $(M,\sigma)$ to the corresponding transaction and send the transaction. 
\end{enumerate}

\noindent Finally, the procedure for the designated auditor is as follows. 
Let $(M,\sigma)$ is generated by specifying $\opk_j$. Run $(\pk_i,\pi)\leftarrow\Open(M,R,\sigma,\osk_j)$. Anyone can check whether $(M,\sigma)$ is generated by the token admitter $i$ by running $\Judge(\opk_j,M,R,\sigma,\pk_i,\pi)$. 

\subsection{Security Analysis} 
Briefly, the security of the proposed scheme is analyzed as follows. 


\begin{itemize}
\item Anonymity: Due to the anonymity of accountable ring signatures, token admitters can issue tokens to addresses anonymously, and anyone can verify it. Only the designated opener can identify the token admitter from the signature. 
Because of the hiding property, adding a commitment $C$ as part of the message does not affect anonymity. Since token submitters send tokens to users on behalf of token admitters, information of token admitters is not revealed from the address. 

\item Auditability: Token admitters have no way to produce untraceable signatures due to the traceability of accountable ring signatures. 
The designated opener can produce a proof of tracing, and due to the tracing soundness of accountable ring signatures, a signature cannot trace to two different signers.

\item Unforgeability: Due to the unforgeability of accountable ring signatures, the address and the flag specified by a token admitter cannot be modified even by the corresponding token submitter because a target address and the flag is contained to a signed message $M$. 

\item Revocability: If two tokens are generated by the same token admitter, then they are linked via $C$ and $(r_{\sf link},\dec)$. Because of the binding property, other token admitters cannot prepare $(r_{\sf link}^\prime,\dec^\prime)$ where $(r_{\sf link}^\prime,\dec^\prime)\neq (r_{\sf link},\dec)$ that ${\sf ComOpen}({\sf crs},C,r^\prime_{\sf link},\dec^\prime)=1$ holds. 

\end{itemize}

We remark that token admitters can generate nonlinked tokens by randomly choosing $C$. However, the designated auditor can identify token admitters even if they generate such nonlinked tokens. For example, a token verifier can ask the designated auditor who the actual token admitter is when the corresponding transaction is not found in the step 4. That is, token admitters have no merit to generate such tokens, and thus we assume that a token verifier can always find the corresponding transaction if tokens are revoked.

\section{Implementation}

In this section, we show our implementation results, in which the Bootle et al. accountable ring signature scheme~\cite{bootle2015short} and the Pedersen commitment scheme~\cite{Pedersen91}. We also employed Curve25519~\cite{Bernstein06} and checked that our scheme is applicable to Bitcoin, Ethereum, and NEM. We used MacBook Pro (2.3 GHz Intel Core i5, 16 GB 2133 MHz LPDDR3) and specified the ring size $N=16$.\footnote{We consider~\cite{monero-ringsize}, which fixes the ring size to 11 in general.} According to this setting, we set $n=4$ and $m=2$ as defined in~\cite{bootle2015short}, where $N=n^m$. 
Then, the signature size is as follows: $(\log_2 N+12)|\mathbb{G}|+\frac{1}{3}(3\log_2 N+12)|\mathbb{Z}_p|\approx 1.4kB$. In this setting, the running times of the $\Sign$ and $\Verify$ algorithms are given in Table~\ref{RunTime_ARS}. We also show the running times of the {\sf Commit}/{\sf ComOpen} algorithms in Table~\ref{RunTime_COM}. From these implementation results, we can say that both cryptographic tools are reasonable in blockchain systems in terms of efficiency.

\begin{table}[h]
\centering
\caption{Running Time of the Bootle et al. Accountable Ring Signature Scheme}\label{RunTime_ARS}
  \begin{tabular}{|c|c|c|} \hline
    Algorithm & Time (ms) \\ \hline \hline
    $\Sign$ & 13.7 \\ \hline
    $\Verify$ & 11.0 \\ \hline
  \end{tabular}
\end{table}

\begin{table}[h]
\centering
\caption{Running Time of the Pedersen Commitment Scheme}\label{RunTime_COM}
  \begin{tabular}{|c|c|c|} \hline
    Algorithm & Time (ms) \\ \hline \hline
    ${\sf Commit}/{\sf ComOpen}$ & 0.0072 \\ \hline
  \end{tabular}
\end{table}

Next, we confirmed that our scheme is applicable in terms of both computational costs and remittance charge for Bitcoin,  Ethereum, and NEM. 
We estimated them using the exchange rate on April 6, 2020. 
We implemented two cases: 

\begin{enumerate}
\item We embedded a ring signature on the transaction directly. Then, one can check trustworthiness or untrustworthiness of a token unless the blockchain system is down. Owing to high availability of  blockchain systems, it can be a merit of this case. 
On the other hand, we needed to embed 1612 bytes in Bitcoin, 1619 bytes in Ethereum, 1616 bytes in NEM, respectively, for a transaction. 
Because the embeddable data size is limited, it is a demerit of this case. 

\item We embedded a hash value of a ring signature (and its messages) and 
a URL of an outside storage (e.g., Google Drive,, DropBox, and so on) or path of \lq\lq hash-of-object" of a decentralized storage (e.g., InterPlanetary File System (IPFS)\footnote{See \url{https://ipfs.io/}.}), that preserves the ring signature. 
If SHA256 is employed as the hash function, then we can drastically reduce the data size (32 bytes) and it can solve the demerit of the first case. However, we needed to additionally require to confirm whether its hash value is the same as that of the embedded one after verifying the ring signature. 
Moreover, we need to assume that the outside storage is also available, and does not collude with other entities. Here, we simply assumed that the outside storage is semi-honest, i.e., it follows the protocol description. 

\end{enumerate}

\noindent\textbf{Bitcoin}: In Bitcoin, Bitcoin script is used to describe transactions. To embed data to a transaction, we can use OP\_RETURN. Because it embeds 80 bytes of data at once, we needed to use 21 transactions for case 1 and 2 transactions for case 2 when Google Drive URL (66 bytes) or IPFS path (48 Bytes) is used. 
This result shows that we needed to consider some meta-information that connect two (or potentially more) transactions. 
One easy way to do this in case 2 is to employ the transaction identity (txid) which is a 32 byte value. 
We embed the URL to transaction 1, and embed the hash value of ring signatures and the txid of transaction 1 to transaction 2. 
If one would like to connect two transactions, it finds transaction 1 via the txid in transaction 2, downloads a ring signature and the corresponding message from either the Google Drive URL or the path of IPFS in transaction 1, checks whether the signature is valid, and also checks whether its hash value is the same as the hash value contained in transaction 2. Basically, this simple way works in case 1. 
Regarding remittance charges, it is changed according to the data size or the conditions of blockchains, and so on. Moreover, the transaction may not be confirmed depending on remittance charges. 
As a reference, we needed to 0.002 BTC (\euro 13.74) per transaction if the transaction is confirmed within 5 min. So, the costs for case 1 and case 2 are 0.042 BTC (\euro 288.54) and 0.004 BTC (\euro 27.48), respectively.
Although our scheme is applicable to Bitcoin in terms of functionality point of view, there is a room for argument on such high remittance charges. 

\medskip
\noindent\textbf{Ethereum}:  In Ethereum, we embedded a token to data space using the Ethereum wallet called MetaMask. 
Because we can use high availability of blockchain systems, and 1 transaction is enough for both cases, case 1 is better than case 2 for Ethereum. 
We needed to pay 46,888 gas for case 1. When we set a gas price of 2 gwei, it was 0.000093776 ETH (\euro 0.015). 
We remark that our experiment was run on the testnet, and the average gas price of mainnet was 14.38 gwei when the experiment was conducted (April 6, 2020). 
But still, this showed that the remittance fee was reasonable; thus, we concluded that our scheme is applicable to Ethereum. 

\medskip
\noindent\textbf{NEM}:  
In NEM, the transaction/prepare-announce API is used for sending a transaction. To embed data to a transaction, we can use the message space. Because it embeds 1024 bytes of data at once, we needed to use 2 transactions for case 1 and 1 transaction for case 2. We can employ txid to connect two transactions as in the case 2 of bitcoin. 
In NEM version 1, the remittance charge is estimated by fee(XEM) = (MessageLength/32 + 1) $\ast$ 0.05(XEM). 
We needed 2.80 XEM (\euro 0.10) in case 1, and in case 2 we needed 0.25 XEM (\euro 0.009) when Google Drive URL (66 bytes) is used and 0.20 XEM (\euro 0.007) when IPFS path (48 bytes) is used, respectively. 
Because we do not have to connect two transactions, case 2 seems better than case 1 for NEM. 
This results showed that the remittance fee was reasonable; thus, we concluded that our scheme is applicable to NEM. 

\section{Conclusion and Future Work}

In this paper, we proposed an anonymous trust-marking scheme on blockchain systems based on accountable ring signatures and commitment, and confirmed  that our scheme is applicable to Bitcoin, Ethereum and NEM considering both computational costs and remittance charges.

Since our system adds/revokes (un)trustworthiness to users (addresses), we further need to consider how to evaluate (un)trustworthiness of users (addresses) outside of our system. As a simple way, the (un)trustworthiness of users (addresses) can be assumed to be judged by external organizations such as an investigative organization, and is shared to token admitters. We leave considering such an evaluation system run outside of our anonymous trust-marking scheme as a future work of this paper. 
We simply assumed that if untrustworthiness is added to an address, then nobody makes a deal to the address. However, as a next step, it needs to more concretely consider how can token verifiers utilize such tokens during or after a cyber incident, and similarly need to more concretely consider how to utilize trustworthiness token. 
Moreover, how to incentive parties to act as admitters, submitters, or auditors, such as financial incentives, should be further discussed when the scheme will be implemented in a practical deployment. Considering a malicious outside storage are also left as future works. 

\medskip
\noindent\textbf{Acknowledgment}: We thank the reviewers of ICBC 2021 for their invaluable comments and suggestions. 
This work was supported by JSPS KAKENHI Grant Number JP19H04107. 

\bibliographystyle{plain}
\bibliography{ref}

\begin{thebibliography}{10}

\bibitem{nemio_announcement1}
Coincheck hack update: Removal of {M}osaic tagging system.
\newblock
  https://medium.com/nemofficial/coincheck-hack-update-removal-of-mosaic-tagging-system-18b4157ff060.
\newblock Accessed : 2018-04-11.

\bibitem{AndroulakiBBCCC18}
Elli Androulaki, Artem Barger, Vita Bortnikov, Christian Cachin, Konstantinos
  Christidis, Angelo~De Caro, David Enyeart, Christopher Ferris, Gennady
  Laventman, Yacov Manevich, Srinivasan Muralidharan, Chet Murthy, Binh Nguyen,
  Manish Sethi, Gari Singh, Keith Smith, Alessandro Sorniotti, Chrysoula
  Stathakopoulou, Marko Vukolic, Sharon~Weed Cocco, and Jason Yellick.
\newblock {Hyperledger Fabric}: a distributed operating system for permissioned
  blockchains.
\newblock In {\em EuroSys}, pages 30:1--30:15, 2018.

\bibitem{AndroulakiCCDET20}
Elli Androulaki, Jan Camenisch, Angelo~De Caro, Maria Dubovitskaya, Kaoutar
  Elkhiyaoui, and Bj{\"{o}}rn Tackmann.
\newblock Privacy-preserving auditable token payments in a permissioned
  blockchain system.
\newblock In {\em {ACM} Conference on Advances in Financial Technologies},
  pages 255--267, 2020.

\bibitem{au2006short}
Man~Ho Au, Sherman~SM Chow, Willy Susilo, and Patrick~P Tsang.
\newblock Short linkable ring signatures revisited.
\newblock In {\em European Public Key Infrastructure Workshop}, pages 101--115.
  Springer, 2006.

\bibitem{0001HS19}
Michael Backes, Lucjan Hanzlik, and Jonas Schneider{-}Bensch.
\newblock Membership privacy for fully dynamic group signatures.
\newblock In Lorenzo Cavallaro, Johannes Kinder, XiaoFeng Wang, and Jonathan
  Katz, editors, {\em {ACM} {CCS}}, pages 2181--2198. {ACM}, 2019.

\bibitem{BarakGIRSVY01}
Boaz Barak, Oded Goldreich, Russell Impagliazzo, Steven Rudich, Amit Sahai,
  Salil~P. Vadhan, and Ke~Yang.
\newblock On the (im)possibility of obfuscating programs.
\newblock In {\em {CRYPTO}}, pages 1--18, 2001.

\bibitem{BellareSZ05}
Mihir Bellare, Haixia Shi, and Chong Zhang.
\newblock Foundations of group signatures: The case of dynamic groups.
\newblock In {\em {CT-RSA}}, pages 136--153, 2005.

\bibitem{Ben-SassonCG0MTV14}
Eli Ben{-}Sasson, Alessandro Chiesa, Christina Garman, Matthew Green, Ian
  Miers, Eran Tromer, and Madars Virza.
\newblock Zerocash: Decentralized anonymous payments from bitcoin.
\newblock In {\em {IEEE} Symposium on Security and Privacy}, pages 459--474,
  2014.

\bibitem{BenderKM09}
Adam Bender, Jonathan Katz, and Ruggero Morselli.
\newblock Ring signatures: Stronger definitions, and constructions without
  random oracles.
\newblock {\em J. Cryptology}, 22(1):114--138, 2009.

\bibitem{BenjumeaCLY08}
Vicente Benjumea, Seung~Geol Choi, Javier L{\'{o}}pez, and Moti Yung.
\newblock Fair traceable multi-group signatures.
\newblock In {\em Financial Cryptography and Data Security}, pages 231--246,
  2008.

\bibitem{Bernstein06}
Daniel~J. Bernstein.
\newblock Curve25519: New {D}iffie-{H}ellman speed records.
\newblock In {\em Public Key Cryptography}, pages 207--228, 2006.

\bibitem{BootleCCGG16}
Jonathan Bootle, Andrea Cerulli, Pyrros Chaidos, Essam Ghadafi, and Jens Groth.
\newblock Foundations of fully dynamic group signatures.
\newblock In {\em Applied Cryptography and Network Security}, pages 117--136,
  2016.

\bibitem{bootle2015short}
Jonathan Bootle, Andrea Cerulli, Pyrros Chaidos, Essam Ghadafi, Jens Groth, and
  Christophe Petit.
\newblock Short accountable ring signatures based on {DDH}.
\newblock In {\em {ESORICS}}, pages 243--265, 2015.

\bibitem{BunzAZB20}
Benedikt B{\"{u}}nz, Shashank Agrawal, Mahdi Zamani, and Dan Boneh.
\newblock Zether: Towards privacy in a smart contract world.
\newblock In {\em Financial Cryptography and Data Security}, pages 423--443,
  2020.

\bibitem{CamenischDD17}
Jan Camenisch, Manu Drijvers, and Maria Dubovitskaya.
\newblock Practical uc-secure delegatable credentials with attributes and their
  application to blockchain.
\newblock In {\em {ACM} {CCS}}, pages 683--699, 2017.

\bibitem{CamenischS97}
Jan Camenisch and Markus Stadler.
\newblock Efficient group signature schemes for large groups (extended
  abstract).
\newblock In {\em {CRYPTO}}, pages 410--424, 1997.

\bibitem{ChaumH91}
David Chaum and Eug{\`{e}}ne van Heyst.
\newblock Group signatures.
\newblock In {\em {EUROCRYPT}}, pages 257--265, 1991.

\bibitem{FauziMMO19}
Prastudy Fauzi, Sarah Meiklejohn, Rebekah Mercer, and Claudio Orlandi.
\newblock Quisquis: {A} new design for anonymous cryptocurrencies.
\newblock In {\em {ASIACRYPT}}, pages 649--678, 2019.

\bibitem{Garman0M16}
Christina Garman, Matthew Green, and Ian Miers.
\newblock Accountable privacy for decentralized anonymous payments.
\newblock In {\em Financial Cryptography and Data Security}, pages 81--98,
  2016.

\bibitem{Ghadafi14}
Essam Ghadafi.
\newblock Efficient distributed tag-based encryption and its application to
  group signatures with efficient distributed traceability.
\newblock In {\em {LATINCRYPT}}, pages 327--347, 2014.

\bibitem{GrothK15}
Jens Groth and Markulf Kohlweiss.
\newblock One-out-of-many proofs: Or how to leak a secret and spend a coin.
\newblock In {\em {EUROCRYPT}}, pages 253--280, 2015.

\bibitem{GrothO07}
Jens Groth and Rafail Ostrovsky.
\newblock Cryptography in the multi-string model.
\newblock In {\em {CRYPTO}}, pages 323--341. Springer, 2007.

\bibitem{KumawatP17}
Sudhakar Kumawat and Souradyuti Paul.
\newblock A new constant-size accountable ring signature scheme without random
  oracles.
\newblock In {\em Inscrypt}, pages 157--179, 2017.

\bibitem{KustersRS20}
Ralf K{\"{u}}sters, Daniel Rausch, and Mike Simon.
\newblock Accountability in a permissioned blockchain: Formal analysis of
  hyperledger fabric.
\newblock {\em {IACR} Cryptology ePrint Archive}, 2020:386, 2020.

\bibitem{LaiZCS16}
Russell W.~F. Lai, Tao Zhang, Sherman S.~M. Chow, and Dominique Schr{\"{o}}der.
\newblock Efficient sanitizable signatures without random oracles.
\newblock In {\em {ESORICS}}, pages 363--380, 2016.

\bibitem{rfc7748}
Adam Langley, Mike Hamburg, and Sean Turner.
\newblock {Elliptic Curves for Security}.
\newblock RFC 7748, January 2016.

\bibitem{LiuWW04}
Joseph~K. Liu, Victor~K. Wei, and Duncan~S. Wong.
\newblock Linkable spontaneous anonymous group signature for ad hoc groups
  (extended abstract).
\newblock In {\em {ACISP}}, pages 325--335, 2004.

\bibitem{Lu0ZL19}
Tingting Lu, Jiangtao Li, Lei Zhang, and Kwok{-}Yan Lam.
\newblock Group signatures with decentralized tracing.
\newblock In {\em Inscrypt}, pages 435--442, 2019.

\bibitem{LuAZ19}
Xingye Lu, Man~Ho Au, and Zhenfei Zhang.
\newblock Raptor: {A} practical lattice-based (linkable) ring signature.
\newblock In {\em Applied Cryptography and Network Security}, pages 110--130,
  2019.

\bibitem{MiersG0R13}
Ian Miers, Christina Garman, Matthew Green, and Aviel~D. Rubin.
\newblock Zerocoin: Anonymous distributed e-cash from bitcoin.
\newblock In {\em {IEEE} Symposium on Security and Privacy}, pages 397--411,
  2013.

\bibitem{NakanishiFW1999}
Toru Nakanishi, Toru Fujiwara, and Hajime Watanabe.
\newblock A linkable group signature and its application to secret voting.
\newblock {\em {JIP}}, 40(7):3085--3096, 1999.

\bibitem{Noether15}
Shen Noether.
\newblock Ring signature confidential transactions for {M}onero.
\newblock {\em {IACR} Cryptology ePrint Archive}, 2015:1098, 2015.

\bibitem{OSS20}
Kazumasa Omote, Asuka Suzuki, and Teppei Sato.
\newblock A new method of assigning trust to user address in bitcoin.
\newblock In {\em The Second {IEEE} International Conference on Blockchain
  Computing and Applications}, pages 1--6, 2020.

\bibitem{Pedersen91}
Torben~P. Pedersen.
\newblock Non-interactive and information-theoretic secure verifiable secret
  sharing.
\newblock In {\em {CRYPTO}}, pages 129--140, 1991.

\bibitem{PoelstraBFMW18}
Andrew Poelstra, Adam Back, Mark Friedenbach, Gregory Maxwell, and Pieter
  Wuille.
\newblock Confidential assets.
\newblock In {\em Financial Cryptography and Data Security}, pages 43--63,
  2018.

\bibitem{rivest2001leak}
Ronald~L Rivest, Adi Shamir, and Yael Tauman.
\newblock How to leak a secret.
\newblock In {\em International Conference on the Theory and Application of
  Cryptology and Information Security}, pages 552--565. Springer, 2001.

\bibitem{SakaiSEHO12}
Yusuke Sakai, Jacob C.~N. Schuldt, Keita Emura, Goichiro Hanaoka, and Kazuo
  Ohta.
\newblock On the security of dynamic group signatures: Preventing signature
  hijacking.
\newblock In {\em Public Key Cryptography}, pages 715--732, 2012.

\bibitem{monero-ringsize}
Riccardo Spagni.
\newblock Monero 0.13.0 \lq\lq {B}eryllium {B}ullet'' release.
\newblock https://web.getmonero.org/tr/2018/10/11/monero-0.13.0-released.html.
\newblock Accessed : 2018-10-11.

\bibitem{SunALY17}
Shifeng Sun, Man~Ho Au, Joseph~K. Liu, and Tsz~Hon Yuen.
\newblock {RingCT} 2.0: {A} compact accumulator-based (linkable ring signature)
  protocol for blockchain cryptocurrency monero.
\newblock In {\em {ESORICS}}, pages 456--474, 2017.

\bibitem{TorresSSLKBAC18}
Wilson Abel~Alberto Torres, Ron Steinfeld, Amin Sakzad, Joseph~K. Liu, Veronika
  Kuchta, Nandita Bhattacharjee, Man~Ho Au, and Jacob Cheng.
\newblock Post-quantum one-time linkable ring signature and application to ring
  confidential transactions in blockchain (lattice {RingCT} v1.0).
\newblock In {\em {ACISP}}, pages 558--576, 2018.

\bibitem{van2013cryptonote}
Nicolas Van~Saberhagen.
\newblock {CryptoNote} v 2.0, 2013.

\bibitem{WangCM19}
Xueli Wang, Yu~Chen, and Xuecheng Ma.
\newblock Adding linkability to ring signatures with one-time signatures.
\newblock In {\em {ISC}}, pages 445--464, 2019.

\bibitem{WatanabeIO0NHK19}
Hiroki Watanabe, Tatsuro Ishida, Shigenori Ohashi, Shigeru Fujimura, Atsushi
  Nakadaira, Kota Hidaka, and Jay Kishigami.
\newblock Enhancing blockchain traceability with {DAG}-based tokens.
\newblock In {\em {IEEE} {Blockchain}}, pages 220--227, 2019.

\bibitem{WesterkampVK18}
Martin Westerkamp, Friedhelm Victor, and Axel Kupper.
\newblock Blockchain-based supply chain traceability: Token recipes model
  manufacturing processes.
\newblock In {\em {IEEE} {Blockchain}}, pages 1595--1602, 2018.

\bibitem{XuY04}
Shouhuai Xu and Moti Yung.
\newblock Accountable ring signatures: {A} smart card approach.
\newblock In {\em {CARDIS}}, pages 271--286, 2004.

\bibitem{Yuen19}
Tsz~Hon Yuen.
\newblock {PAChain}: Private, authenticated and auditable consortium
  blockchain.
\newblock In {\em Cryptology and Network Security}, pages 214--234, 2019.

\bibitem{YuenSLAEZG20}
Tsz~Hon Yuen, Shifeng Sun, Joseph~K. Liu, Man~Ho Au, Muhammed~F. Esgin,
  Qingzhao Zhang, and Dawu Gu.
\newblock {RingCT} 3.0 for blockchain confidential transaction: Shorter size
  and stronger security.
\newblock In {\em Financial Cryptography and Data Security}, pages 464--483,
  2020.

\bibitem{ZhangLLYAW19}
Lingyue Zhang, Huilin Li, Yannan Li, Yong Yu, Man~Ho Au, and Baocang Wang.
\newblock An efficient linkable group signature for payer tracing in anonymous
  cryptocurrencies.
\newblock {\em Future Gener. Comput. Syst.}, 101:29--38, 2019.

\bibitem{ZhangLSKY19}
Xinyu Zhang, Joseph~K. Liu, Ron Steinfeld, Veronika Kuchta, and Jiangshan Yu.
\newblock Revocable and linkable ring signature.
\newblock In {\em Inscrypt}, pages 3--27, 2019.

\end{thebibliography}

\end{document}